%% file: main.tex
\documentclass[nonacm]{acmart}
\AtBeginDocument{%
  \providecommand\BibTeX{{%
    \normalfont B\kern-0.5em{\scshape i\kern-0.25em b}\kern-0.8em\TeX}}}

\usepackage{listings}
\usepackage{xcolor}
\usepackage{algorithm}
\usepackage{algpseudocode}

\begin{document}

\title{Flexible Vector Integration in Embedded RISC-V SoCs for End-to-End CNN Inference Acceleration}

\author{Dmitri Lyalikov}
\email{Dlyalikov01@manhattan.edu}
\orcid{0009-0007-6024-0734}
\affiliation{
  \institution{Manhattan College}
  \streetaddress{4513 Manhattan College Pkwy}
  \city{Riverdale}
  \state{New York}
  \country{USA}
  \postcode{10471}
}

\renewcommand{\shortauthors}{Lyalikov, et al.}

\begin{abstract}
The emergence of heterogeneity and domain-specific architectures targeting deep learning inference show great potential for enabling the deployment of modern CNNs on resource-constrained embedded platforms. A significant development is the diversification of custom hardware solely targeting the most expensive parts of CNNs. DLAs (deep learning accelerators) and NPUs (neural processing units), among others, can overcome the approaching limits of traditional silicon scaling and provide a solution to the power/performance tradeoff within embedded SoCs. An accelerator-rich SoC hoists the bulk computation and data stream away from the scalar control CPU and onto optimized domain-specific accelerators. Efficient DSA utilization requires proper system integration and a compilation/execution model for balanced execution in these heterogeneous architectures. Without a system-specific hardware-software codesign, a  problem of a disjointly balanced execution model can arise, where a large portion of processing time is spent on CPU fallback computation for operations not supported by the custom architectures. Therefore, there is a critical need for proper system integration and an efficient compilation/execution model for balanced execution in these heterogeneous architectures. Building on recent access to open-source hardware, we explore the software-hardware codesign process of integrating and optimizing the end-to-end dataflow of deep learning accelerator into a RISC-V System-on-Chip. This work highlights the hardware integration challenges for efficiently placing these units within the memory hierarchy and correct proximity to other execution blocks. We also experimentally verify performance bottlenecks in CNN execution and pre/post-processing at runtime, where previous attention has generally been given to accelerator speedup alone. This work takes advantage of the ratification of the RISC-V Vector 1.0 extension and demonstrates its potential as a flexible target within a well-suited cache hierarchy scheme to reduce pre-processing bottlenecks and CPU fallback processes. Our results show up to a 9x speedup of image pre-processing and YOLOv3 fallback layer execution by up to 3x compared to CPU. We demonstrate RVV-1.0 in exposing a flexible programming model that can enable a balanced computation and memory footprint on accelerator-rich embedded SoCs supporting modern deep-learning dataflows while consuming less power than traditional parallel execution platforms.
\end{abstract}



\maketitle

\section{Introduction}
The rapidly growing demand for deployment of compute-intensive CNN inference on resource-constrained embedded systems such as automotive driver assistance systems (ADAS) and autonomous drones has pushed industry and academia to revolutionize the traditional silicon scaling paradigm to push far beyond the limits of Moore's law. SoC (System-on-Chip) designs now favor a more heterogeneous architecture model to support the execution and data requirements of these application models. More significant areas on an embedded SoC are now being allocated to domain-specific accelerators. This exposes a software/hardware problem of how to efficiently utilize DSA resources to maximally realize their potential. An efficient system-level integration requires a deep understanding of the target architecture(s) and provides meaningful tools and frameworks that can support them. To add to that, there must be a thorough examination of the application dataflow, its execution patterns, and memory requirements. These challenges have presented themselves to the entire stack of the SoC design spectrum, from hardware architecture through compiler technology and up to high-level software implementation.

Convolutional Neural Networks are some of the most expensive in deep learning. Their heavy memory footprint and large convolution and matrix operations have been prohibitively hard to execute in real-time on traditional scalar CPUs without support from a power-hungry graphics processing unit (GPU). There has been an enormous push in recent years from semiconductor companies to implement specialized hardware units that optimize the computation patterns and memory accesses of CNN inference. These dedicated modules are typically designed to perform convolution and matrix operations very efficiently, which can absorb more than 90\% of the CNN operations \cite{Survey-DLACHEN2020264}. CNN hardware can generally be categorized by their generality in implementation, from general-purpose hardware capable of CNN dataflows, dedicated architecture that are targetting CNNs, and nueromorphic hardware, that is inspired by the functional model of the human brain \cite{DLA-Comp-survey9222299}.

Alongside the rise of heterogeneous computing architecture, the emerging prominence of the RISC-V ISA has shown to be a promising enabling technology, with RISC-V AI SoCs expected to have a CAGR for units of 73.6\% by the year 2027. 
This royalty free instruction set architecture exposes a high degree of freedom in its base ISA implementation, and is coupled with a configurable set of application specific extensions that can be removed, included, or added to via custom instructions. These luxuries allow architects and design teams to have more freedom in implementation, and grants more resource availability to creating leaner, more efficient implementations. We are already seeing the introduction of RISC-V AI SoCs trickling down into accessible commercial domains. The Beagle-V ahead board 
has been put to market by Texas Instruments as of 2023, and includes the TH1520 Quad-Core RISC-V Processor and features a 4 TOPs NPU (Neural Processing Unit) with full support for the Linux kernel. The Kendryte K-series are devices brought to market under the edge computing term, including RISC-V Cores as the platform of choice to support real time deep learning and computer vision workloads. Microchip Technology has introduced a series of Polarfire FPGA platforms that include a RISC-V SoC, and taking advantage of the Asymmetric Multiprocessing Mode (AMP) the ISA allows, provides the ability to run a Linux OS and RTOS simultaneously. 
The Milk-V company has embraced this architecture and offer high performance AI, image processing, and OS capabilities in numerous devices that share a similar form factor to the Raspberry Pi Pico, Pi 4B, and Pi Compute Module. 
As RISC-V continues to grow, the ecosystem, tooling, and availability will migrate and create a well suited solution to platforms supporting modern CNN applications. 

To support the increasing range and diversified architectures in an embedded SoC, it is critical that compiler technology and the invocation/configuration model for these designs keeps up with their advances. There is a need for hardware aware IR that can optimize node and block level execution kernels and find the correct execution unit to assign to. In the event that the deep-learning accelerator does not support an execution kernel, the hardware-software optimization problem evolves into how to efficiently deal with fall back operations on less prioritized hardware. 

In this paper, we provide a comprehensive analysis of the integration challenges of modern embedded SoCs for CNN inference. We do this by evaluating a wide design spectrum, including micro-architecture, system-level design, compiler technology, deep learning frameworks and high level software design. This does not provide an exhaustive dissection of one or more, rather a specific description of how each area relates to problems and design considerations in terms of power and performance that can be generalized. We also evaluate the RISC-V Vector extension as a suitable target to minimize latency in CPU fallback operations and image pre/processing. To the best of our knowledge, this is the first paper that provides a detailed synthesis and practical implementation on software/hardware RISC-V vector integration for modern CNN applications. More specifically, this paper contributes the following:
\begin{itemize}
    \item We dissect the current state of deep learning acceleration technology. This includes accelerator architectures and the compilers that support them. We elaborate the problem of diversifying architectures and heterogeneity in SoC and the requirement for DL compilers and software stack to efficiently support them
    \item We evaluate the YOLOv3 CNN which has been ported to a RISC-V SoC including a DLA. We take a system level approach and identify performance bottlenecks that contribute to latency and power inefficiency. We provide a comprehensive analysis of the deep-learning model architecture and explore the contributing factors to inefficiencies in deploying accelerated CNNs.
    \item We provide a design example of integrating a Vector Coprocessor into a DLA integrated system and show the memory system considerations to maximize throughput in the Chipyard framework \cite{Chipyard:9099108}. We provide a detailed description of vector mapping of CPU fallback procedures and layers and document the speedup potential. We contribute VecBoost, an open-source vector library of common fallback operations that supports RVV 1.0.
    \item We give several insights into future development of embedded CNN SoCs, and highlight the RISC-V Vector extension and its need for suitable auto-vectorization tools, to provide easier software integration into existing DL deployment stacks.
\end{itemize}

The rest of this paper is organized as follows. Section 2 provides the background of deep learning execution and system design. Section 3 describes system level design considerations. Section 4 discusses the architecture and deployment details of the YOLOv3 CNN on NVDLA, and the performance results of each layer. Section 5 describes the agile design methodology used to evaluate and simulate the SoC architecture and the results on the base system. Section 6 elaborates parallel processing solutions for optimizing YOLOv3 and CNN bottlenecks and compares vector processors with traditional options. Section 7 details the vector integration into the base SoC and tuning the cache parameters for optimal results. It also provides a comprehensive example of the RVV 1.0 vectorization process and compares speedup results with base CPU performance.

\section{Background and Motivations}
In this section, we elaborate the resource and computational demands of modern CNNs and describe the current hardware that can support it. This is not meant to be exhaustive, but to provide a contextual overview for deep learning integration. 
\subsection{CNN Execution and Dataflow}
While much focus in the CNN dataflow is spent between first layer input activations until the very last layer, a significant amount of work appears in pre-processing and configuration steps before the CNN inference even begins. Therefore, it is important to understand the behavior from the time an image frame arrives in main system memory to the point it has been post-processed and the useful information extracted in CNN inference is available. In an ideal system, the CPU subsystem within an SoC has the role of configuration and management of dataflows streaming in and out of DLA(s) and is generally isolated from computation. Many times, this is not the case. \cite{gray2016grvi}
\subsection{Deep Learning Hardware}
DNN accelerators can generally be organized under three categories, neuromorphic accelerators, which are inspired by the execution patterns of the human brain, dataflow accelerators, and configurable deep learning accelerators. Dataflow accelerators are application specific architectures that take dataflows represented in a loop nest form (seen in Equation \ref{eq:loopnest}) and use loop ordering of row-order form  and spatial unrolling to map the execution into an array of small-sub accelerators \cite{DLA-dataflow9407116}. This process flattens the execution graph and allows each execution unit (kernel) to be mapped independently to an array of processing elements. 
Loop nest form:
\begin{equation}\label{eq:loopnest}
    \sum_{i = L_1}^{U_1} \sum_{j = L_2}^{U_2} \sum_{k = L_3}^{U_3} \ldots \sum_{n = L_n}^{U_n}
\end{equation}
Where 
\begin{align*}
i, j, k &: \text{Loop variable for the first, second, third dimensions.}  \\ 
    L_1, L_2, L_3 &: \text{Lower bounds for each loop variable.}  \\ 
    U_1, U_2, U_3 &: \text{Upper bounds for each loop variable} \\
\end{align*}
Spatial unrolling: 
\begin{equation}\label{eq:spatialunrolling}
    \sum_{i=0}^{N-1} \sum_{j=0}^{M-1} \ldots \sum_{k=0}^{P-1} f(i, j, \ldots, k)
\end{equation}
Where
\begin{align*}
    N, M, P &: \text{Upper bounds for each dimension} \\
    f(i, j, ..., k) &: \text{Computation or operation at each iteration point} \\
\end{align*}
Row-order form:
\begin{equation}\label{eq:rowunrolling}
    \sum_{i=0}^{n-1}\sum_{j=0}^{m-1}
\end{equation}
Where
\begin{align*}
    n &: \text{Number of rows} \\
    m &: \text{Number of columns}
\end{align*}

\begin{figure}[ht]
  \centering
  \includegraphics[width=\linewidth]{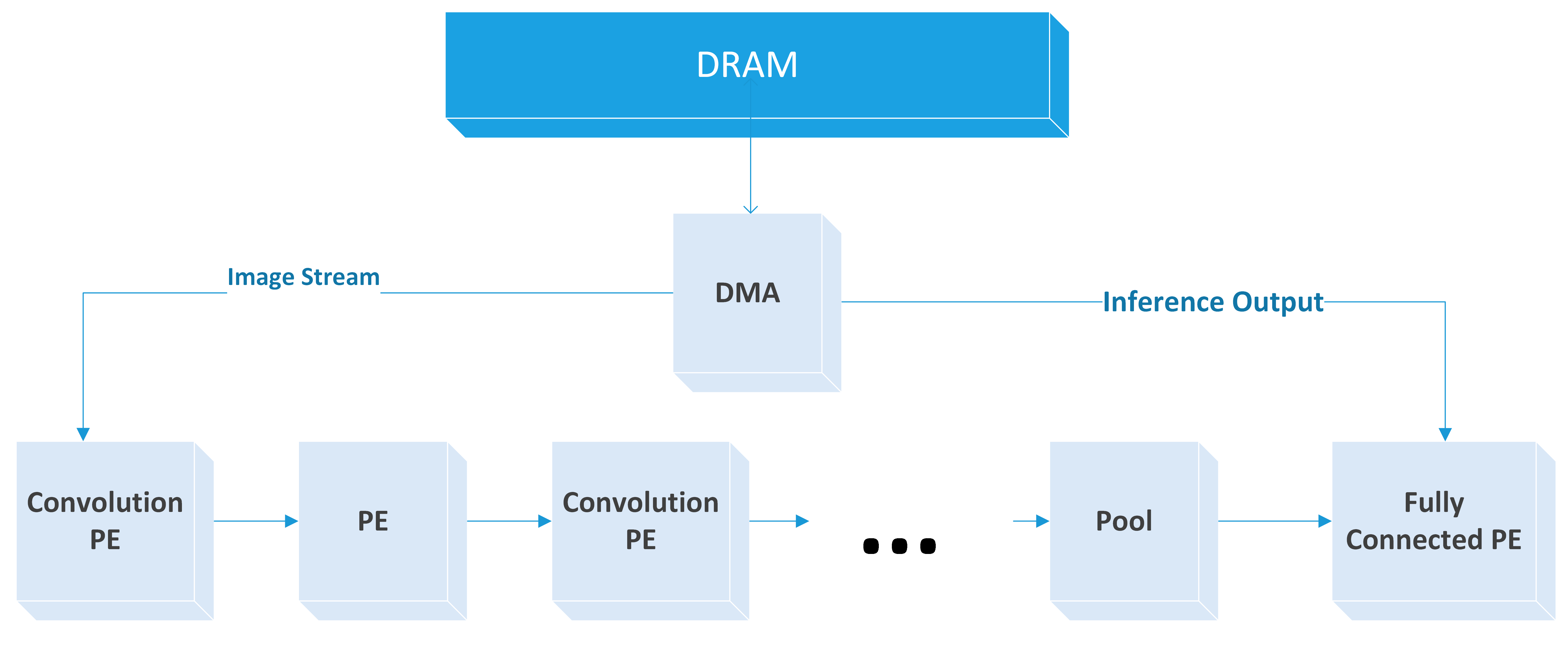}
  \caption{Streaming DLA Architecture}
  \Description{NPU}
\end{figure}

Configurable deep learning accelerators are programmable devices that emphasize efficiency on the most expensive operations within CNNs: convolution, pooling, matrix multiplication, and activation functions. Some examples include Neural Processors (NPU), and Systolic Array generators such as Gemmini \cite{Gemmini9586216}. Neural Processors are essentially an array of processing elements, each of which perform the basic operations of a neuron multiply-accumulate and sigmoid activation \cite{surveyarchCHEN2020264}. In general, all architectures provide a parallel, and highly efficient way of computing the basic blocks of CNNs:

Convolution:
\begin{equation}\label{eq:conv}
    (I * K)(x, y) = \sum_{i=-\infty}^{\infty}\sum_{j=-\infty}^{\infty} I(x-i, y-j) \cdot K(i, j)
\end{equation}

\begin{align*}
    I &: \text{Input feature map} \\
    K &: \text{Convolutional kernel} \\
    (I * K)(x,y) &: \text{Value at position x, y in the output feature}
\end{align*}
Activation (ReLU) Rectified Linear Unit is the most common activation function in CNNs, introducing non-linearity to the model:
\begin{equation}\label{eq:relu}
\text{ReLU}(x) = \max(0, x)
\end{equation}

Max Pooling to downsample feature maps after convolution. This reduces spatial dimensions:
\begin{equation}\label{eq:maxpool}
    \text{MaxPooling}(x, y) = \max\left(\begin{array}{cccc} P(x, y) & P(x+1, y) \\ P(x, y+1) & P(x+1, y+1) \end{array}\right)
\end{equation}
Where
\begin{align*}
    P &: \text{The input feature map and MaxPooling(x, y) is output with 2x2 window}
\end{align*}

\begin{figure}[ht]
  \centering
  \includegraphics[width=\linewidth]{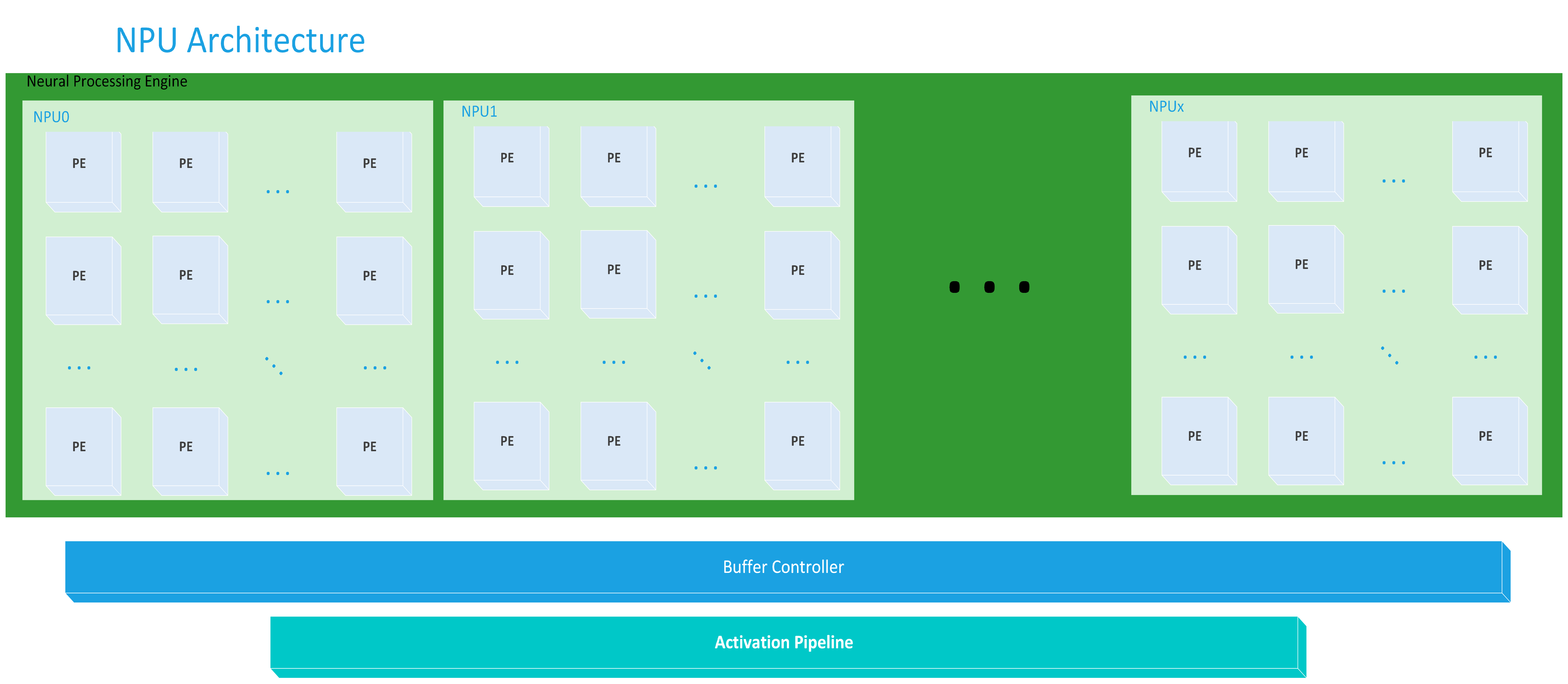}
  \caption{Neural Processor DLA Architecture}
  \Description{NPU}
\end{figure}

\section{CNN SoC Integration Challenges}
This section elaborates the challenges of deploying CNN models into efficient execution models on specialized, domain specific architectures. 
\subsection{Model Compilation}
A big challenge in deploying deep CNNs on DLAs is the problem of efficiently mapping the existing model to the hardware platform \cite{compilerDBLP:journals/corr/abs-2002-03794}. Also, the vast majority of of deep and convolutional neural network models are defined using frameworks. It is the role of the DL compiler to ingest a model specified from TensorFlow, Caffe, or Onnx and create a loadable binary that can be executed. Most DL compilers such as TensorFlow XLA and TVM employ a Front IR (intermediate representation) and a back IR (\cite{TVMDBLP:journals/corr/abs-1802-04799}, \cite{XLA10.1145/3377811.3380434}). The Front IR represents the entire neural network as a graph. It makes decisions for precision of each layer and performs node and block level optimizations including dead code elimination and operator sinking and fusion to simplify the algebraic representation. This is passed to the backend which performs loop oriented optimizations (see equations \ref{eq:loopnest}, \ref{eq:rowunrolling}, \ref{eq:spatialunrolling} above), memory allocation, and assignment to hardware functional units. Here, the compiler backend makes a decision based on its knowledge of the hardware target where each executional unit will be assigned. In some cases, if a fallback mechanism does not exist for a given data precision, layer, or operation on the accelerator target, it will be mapped to a CPU, with clearly less efficiency than expected, or in the worst case, not compile at all. If successful, the Back IR is uses JIT (Just in Time) compilation for code generation, which now is an executable loadable binary. In some cases, this binary can simply be executed within the CPU subsystem, if the accelerator, NPU, or DLA is embedded within the processor ABI. In other situations, where the CPU is acting as a configuration manager to the DLA, a driver and runtime environment may be applicable to load the binary from system memory into a deep learning accelerator's instruction space. 
\subsection{Memory Hierarchy}
In order to make a decision on accelerator integration, it is critical to understand the memory and computation requirements of the intended application \cite{coarse10.1145/3358198}
Accelerators and co-processors within a System-on-Chip can generally be categorized under two groups: tightly coupled and loosely coupled, based on placement within the memory hierarchy and proximity to the scalar control CPU system \cite{coupling7167228}. The first is a tightly-coupled accelerator,  which operates very closely to the scalar CPU and has very little overhead in invocation. TCAs are given access as a participant to the coherent L1 and L2 cache hierarchy and are generally very efficient in piece-wise, procedure and function level acceleration. Some examples of TCAs are floating point units and SHA3 coprocessors, whose operations require few memory accesses and are most compute-bound. They perform very well for offloading repetitive computations in an application where the control CPU has very tight access to its execution and configuration.

\begin{figure}[ht]
  \centering
  \includegraphics[width=\linewidth]{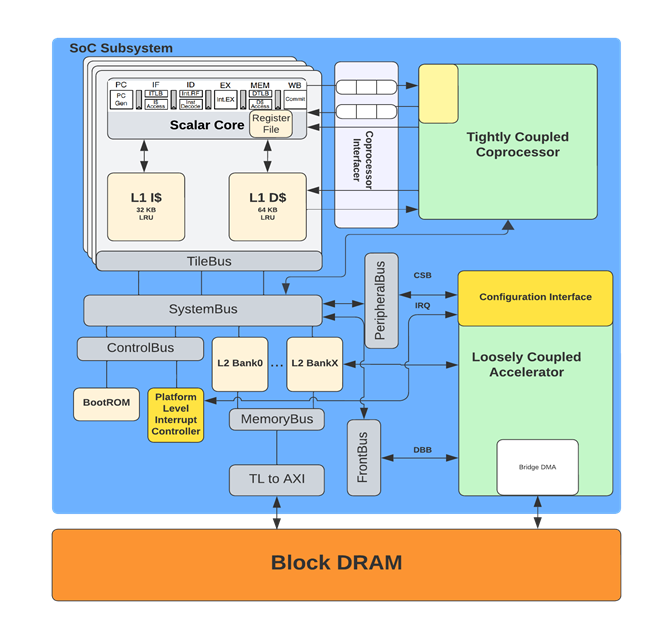}
  \caption{TCA and LCA Architecture}
  \Description{NPU}
\end{figure}

Loosely coupled accelerators operate as independent modules within an SoC. They are generally removed from the cache hierarchy (or only access to the L2 cache) and have direct access to main memory serviced by a DMA controller to reduce the overhead of memory accesses. They may have an internal scratchpad memory and typically take up more silicon area compared to a TCA. LCAs are typically bound by memory access and are suitable for streaming applications such as image/video processing where a frame should be sent directly to an accelerator unit upon arrival. A configuration interface is exposed to the control CPU via a allocated register block in MMIO address space and control status can be advertised via an interrupt controller. Program and execution flow is generally larger compared to a TCA. LCAs can be configured by the CPU for a specific mode of operation, and generally run independently without constant management or computation sharing from the CPU. This frees the CPU from the process. CNN and Deep Learning accelerators are typically designed around the profile of a LCA due to their streaming memory footprints and complex execution behavior. They are typically comprised of multiple block arrays for matrix and convolution calculations.

\section{Design and Implementation}
In order to perform a design space exploration for accelerator rich SoC, the Chipyard \cite{Chipyard:9099108} framework was used. Chipyard provides a set of open-source parameterizable hardware generators and simulation infrastructure exposed through the Chisel HDL \cite{Chisel6241660}.  It provides a collection of reusable components, design methodologies, and infrastructure for designing and implementing custom hardware from scratch \cite{Chipyard:9099108}. Chipyard is built on top of the RISCV64G Rocket Chip generator \cite{RocketAsanović:EECS-2016-17} as its base CPU inside of the SoC subsystem known as a Tile. It offers a wide range of features to support the complete chip design flow, including processor cores, memory systems, peripherals, accelerators and interconnect fabrics.
This work built on the recent successful integration of NVDLA into a Chipyard/FireSim environment for cycle accurate system simulation \cite{NVDLA-FireSim:DBLP:journals/corr/abs-1903-06495}, \cite{karandikar-firesim-isca18}. This platform enabled the fine grained control over the SoC subsystem components through a parameterized system definition and allowed a system level evaluation of the software-hardware codesign process for optimizing around a CNN dataflow.
\subsection{SoC System}
To simulate a heterogeneous SoC in Chipyard, the Chisel elaboration of the design could be compiled for a number of environments, namely Verilator \cite{Verilator9399725} and FireSim. 
Using the Chisel hardware generators Chipyard provides, an evaluation SoC tile was parameterized. Target software was tested against the base DarkNet-YOLOv3 implementation for functional and ABI correctness using the Spike ISA \cite{SpikeHERDT2021102135} simulator, which provides a software-based basic emulator of a RISCVx execution. 

\begin{table}[h]     \centering     \caption{Base SoC Parameterization}     \label{tab:base}     \small    \setlength{\tabcolsep}{6pt} 
\begin{tabular}{|p{4cm}| p{5cm}|} 
\toprule         
\textbf{Unit} & \textbf{Description}  \\     
\hline
\midrule         
\hline
CPU & Quad-Core Rocket 5-stage RVGMAFD  \\  
\hline
L1 I/D\$ Cache & Private 16/16 KiB, 4-way, 64B  \\   
\hline
L2 D\$ Cache & 4MiB, 16-way, 64B LRU  \\         
\hline
DRAM & 2GB \\  
\hline
NVDLA & Large: 2048 INT8 MACs, 512 Kib buffer \\
\hline
\bottomrule     
\end{tabular} 
\end{table}
\subsection{NVDLA}
NVDLA (Nvidia Deep Learning Accelerator) is a recently open sourced hardware IP that is a modular accelerator targeting deep learning inference. Functionally, it operates as a loosely coupled accelerator and accesses main memory over a synchronous, high-speed data bus. It exposes itself to the CPU system as an MMIO device with configuration space and interrupt interface for advertising task completion. The internal architecure includes a Winograd \cite{WinogradDBLP:journals/corr/abs-2201-10369} and Batch convolution array, and a designated space for storing weights. It also has pooling, activation, and reshape engines. The programming and control model is accessed through Nvidia's User Mode and Kernel Mode drivers, which facilitate loading NVDLA execution graphs, starting execution, and managing state at runtime.  
\subsection{YOLOv3 Benchmark}
The YOLOv3 object detection CNN \cite{redmon2018yolov3} was chosen as a suitable benchmark inference workload because of its popularity in modern real time computer vision applications and the great superiority in accuracy and speed \cite{yolo9182679}. This model is based on the 53-layer DarkNet framework \cite{darknet13} which is written in the C language. The YOLO architecture is a completely convolutional model, meaning it does not use a fully connected layer for classification which is in contrast to other CNN architectures such as ResNet or VGG-16.  This enables the model to operate on inputs of an arbitrary size as well as reduce the number of parameters of the model making it simpler and smaller. This comes with the trade off of a more complex loss function. The YOLOv3 loss function for each box prediction is composed of several terms:
1. Coordinate Loss:

\[
\lambda_{\text{coord}} \sum_{i=0}^{S^2} \sum_{j=0}^{B} \mathbb{1}_{ij}^{\text{obj}} \left[(x_i - \hat{x}_i)^2 + (y_i - \hat{y}_i)^2\right] + \lambda_{\text{coord}} \sum_{i=0}^{S^2} \sum_{j=0}^{B} \mathbb{1}_{ij}^{\text{obj}} \left[(\sqrt{w_i} - \sqrt{\hat{w}_i})^2 + (\sqrt{h_i} - \sqrt{\hat{h}_i})^2\right]
\]\label{eq:coord}
Here, $\lambda_{\text{coord}}$ is the coefficient for the coordinate loss, $S$ is the grid size, $B$ is the number of predicted bounding boxes per cell, and $\mathbb{1}_{ij}^{\text{obj}}$\label{eq:obj} is an indicator function that equals 1 if the bounding box $j$ in cell $i$ is responsible for the prediction. The terms in the summation calculate the squared differences between predicted and true bounding box coordinates.
2. Classification Loss:
\[
\sum_{i=0}^{S^2} \sum_{j=0}^{B} \mathbb{1}_{ij}^{\text{obj}} (C_i - \hat{C}_i)^2
\]\label{eq:class}
This term computes the squared difference in class probabilities ($C_i$ and $\hat{C}_i$) for the predicted and true classes of the bounding box.
3. Objectness Loss:
\[
\lambda_{\text{noobj}} \sum_{i=0}^{S^2} \sum_{j=0}^{B} \mathbb{1}_{ij}^{\text{noobj}} (C_i - \hat{C}_i)^2
\]\label{eq:objloss}
The objectness loss term is scaled by $\lambda_{\text{noobj}}$ and calculates the squared difference in class probabilities for cells that do not contain objects, indicated by $\mathbb{1}_{ij}^{\text{noobj}}$.Each of these terms contributes to the total YOLOv3 loss. Using image input resolution augmentation in 10 steps, the model can generalize on a large range of input sizes. These designs make YOLOv3 perform well with a mAP (mean average precision) score of 37 over a range of standard input resolutions (320x320, 416x416, and 618x618).

The output from the YOLOv3 forward inference includes a set of bounding boxes with an associated objectness score. The YOLOv3 objectness score \(C\) and dimension predictions \(t_x\), \(t_y\), \(t_h\), and \(t_w\) are defined as:
\(C\) (objectness score) is a value between 0 and 1, representing the likelihood of an object being present in a particular bounding box.
\(t_x\) and \(t_y\) are the predicted offsets for the center of the bounding box.
\(t_h\) and \(t_w\) are the predicted logarithmic transformations of the height and width of the bounding box. In post-processing, Non-max suppression (NMS) is used to discard bounding boxes that other overlapping it of which have a \(C\) score higher than the threshold value.

Farschi et al. \cite{NVDLA-FireSim:DBLP:journals/corr/abs-1903-06495} integrated the DarkNet-YOLOv3 model into the NVDLA-Firesim environment. This work focused on the porting of the model to the loosely-coupled NVDLA architecture and optimizing the cache performance during inference time. That work showed that for a 416x416 input image, the INT8 inference time took 133 ms and 66 ms of that time was spent on CPU processing of layers not supported by the NVDLA. Looking deeper into the DarkNet implementation, Table \pageref{tab:YoloMapping} elaborates each layer and its eventual mapping to its respective computation unit. 

At run time, within a simulated BuildRoot Linux system booted on the Base SoC, the forward inference process is invoked by passing in `yolov3-odla.cfg' which at a high level describes the model to be executed. This included a 19-layer configuration where ODLA layers loaded additional sub-graphs that had YOLOv3 layers mapped to NVDLA. It is important to note that Split, Converter, and loss/IoU, upsampling calculation was kept to the CPU. The ODLA  layers and sub-layers were mapped to NVDLA. These sub-layers were mapped to the expensive convolutional layers. 
\begin{table}[ht]
    \centering 
    \caption{DarkNet-YOLOv3 Layer Mapping to Rocket-NVDLA SoC} 
    \label{tab:YoloMapping}  
    \begin{tabular}{cccc} 
        \toprule 
        \textbf{Layer} & \textbf{Description} &  \textbf{Assigned Unit} & \textbf{Execution Time(ms)} \\ 
        \midrule  
        Converter & int<->fp32, nchw<->fd & CPU  & 4.6 \\ 
        ODLA::Subgraph0 & Load/Ex. sublayers to NVDLA & NVDLA &  24.5  \\ 
        Split & Resize tensor params & CPU & .2 \\ 
        Upsample ODLA & Upsample tensor & CPU & 10.8 \\
        Split & Resize tensor params & CPU  & .2 \\ 
        Converter & int<->fp32, nchw<->fd &  CPU & 4.8 \\
        YOLO & IoU and Cost Calculation & CPU & 7.97\\
        Split & Resize tensor params & CPU & .2 \\
        ODLA::Subgraph1 & Load/Ex. sublayers to NVDLA & NVDLA & 23.3 \\
        Split & Resize tensor params & CPU & .2 \\
        Converter & int<->fp32, nchw<->fd & CPU & 5.3 \\
        YOLO & IoU and Cost Calculation & CPU & 7.81\\
        Split & Resize tensor params & CPU & .2 \\
        Upsample ODLA & Upsample tensor & CPU & 10.8 \\
        Split & Resize tensor params & CPU & .2 \\
        ODLA::Subgraph2 & Load/Ex. sublayers to NVDLA &  NVDLA  & 20.0\\
        Split & Resize tensor params & CPU & .2 \\
        Converter & int<->fp32, nchw<->fd & CPU & 4.3 \\
        YOLO & IoU and Cost Calculation & CPU & 7.64\\
        \bottomrule 
    \end{tabular} 
\end{table} 
\subsection{Image/Frame Preprocessing}
A section of performance that was not accounted for in initial evaluation metrics was the time and complexity of image/video preparation and pre-processing. It is imperative system level design does not pay attention to inference speed alone, while disregarding the total streaming dataflow behavior. This can give unreasonable dis-balance in the CNN pipeline, which while a CNN may have the capability to support a high frame rate, the end application will hold see much worse. For this reason, we evaluated the entire process in  from image (.JPG in this case) to feeding into the first layer of YOLOv3.
\begin{figure}\label{fig:imgproc}
  \centering
  \includegraphics[width=\linewidth]{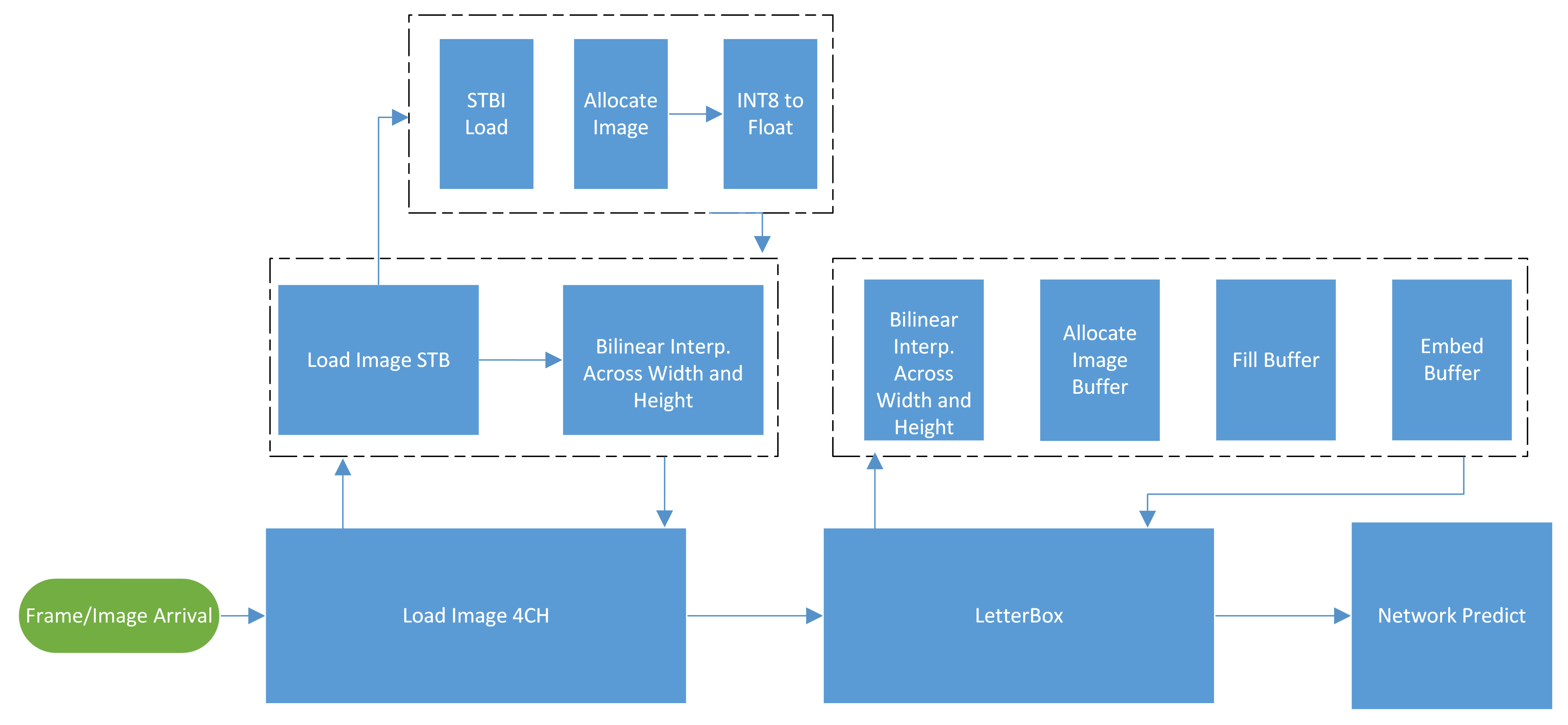}
  \caption{DarkNet Image Pre-Processing Pipeline}
  \Description{DarkNet Image Pre-Processing}
\end{figure}

Many of these operations, written in C and built on the popular STB-I image processing library, were targeting the CPU and performing simple load/store, reformating, and data conversion operations across the matrices. On FireSim Simulation with 100Mhz clocking of the quad-core RISC-V Rocket system, the image preprocessing work took 19.2ms, 27.2ms, and 36.5ms for small, standard, and large image sizes, respectively. These results effectively reduce the expected frame rate by 18\%, with 163 ms end-to-end streaming inference latency for a standard image.

\section{Parallel Computation Models}
The CPU fallback operations and preprocessing procedures that were discovered are not unique. They are mostly simple operations that are operated across the entire dimensions of the input tensor (Height x Width x Channels). For example, the DarkNet-YOLOv3 conversion layer comprises of two distinct procedures: first the conversion from a feature map to NCHW, the typical format of inter-layer tensor passes within a CNN, and second, INT8 to Float32 conversion and vice versa since the CPU and NVDLA operate on different precision. The shared behavior in these operations is that they comprise of triple-nested loops over the Height, Width, and Channels of the input tensor, and perform simple computations or data rearrangements within the inner-most loop, without outer loop dependencies or complex branching. Thus, these functions expose a potential for efficient parallelism. Farschi et al. \cite{NVDLA-FireSim:DBLP:journals/corr/abs-1903-06495} uses the OpenMP pragmas to allow multiprocessing of the inner-loop kernels across the 4 Rocket cores, which helps with a CPU implementation.

There are many existing solutions and hardware approaches to target ultra-parallel execution already. Table \ref{tab:imphw} enumerates the most common platforms and some of their strengths and constraints. The existing CPU implementation the simplest, but takes away the benefit in the accelerated SoC of CPU decoupling from execution. It is also clearly the least efficient, as the memory bottleneck of scalar loads does not lend itself well to streaming data in and out of a L1 data cache. The best alternative on resource rich systems is GPU offloading, which are highly efficient in these types of computations. DarkNet-YOLOv3 framework and most deep-learning models have direct compatibility with streamlined execution to a graphics unit. CUDA and cudaNN are highly optimized tools for mapping parallel workloads onto Nvidia GPUs. They are even capable and often used for entire deep-learning inference and training on large systems. However, the power consumption and area consumption has traditionally been exceedingly large \cite{GPUpower8782524} 
and therefore evading much use in resource constrained embedded devices where a power budge is generally prohibitive of GPUs.

FPGA implementations are also highly efficient platforms for mapping execution onto and generally outperform GPUs in power efficiency for similar image processing workloads \cite{FPGA-power9307865}. There are growing developments in HLS (High Level Software) that can take programming models and functional kernels and map them onto FPGA resources efficiently. Tools like `esp4ml` and the agile SoC design framework, `esp` \cite{espMantovani_2020} can even ingest entire deep-learning models from popular frameworks such as Keras, TensorFlow, and PyTorch and effectively generate a digital design mapping the entire model on FPGA. These tools greatly simplify the long standing design time of implementing an FPGA design in HDL such as Verilog/SystemVerilog as we see a shift towards higher levels of abstraction in describing FPGA applications \cite{deeplearningfpga10.1145/3613963}. However, the ultra-flexible FPGA designs are limited by the amount of CLB (Configurable Logic Blocks) and other resources they can offer, and are generally clocked at rates of hundreds of Megahertz. With the general trend in CNN and other deep-learning models growing in size and complexity \cite{CNNgrowtharticle}, it can be expected that an embedded SoC design integrating FPGAs will have resource constraints on larger, more efficient models. Alongside this, modern FPGAs are complex devices with very high cost, ranging from hundreds of dollars to many thousands of dollars for resource rich platforms. This is generally prohibitive within the field of IoT and embedded devices where net cost is a fraction of this price.

ASICs (Application Specific Integrated Circuits) are fixed function hardware circuits. Modern ASICs can be designed also with HLS and are notorious for being ultra efficient and low footprint. However, their lack of flexibility makes ASIC integration to solve a wide range of CPU fallback operations from different models and targets unlikely.  

The vector architecture is specified as an extension to the RISC-V base ISA. It is similar to the SIMD programming model, in that data elements of a given size are operated on concurrently, per instruction, in contrast to the sequential scalar CPU. The elegant feature Vector ISAs boast over SIMD is that they push the data width and vector size to registers instead of being encoded into a field within instructions like SIMD. This overcomes the long standing problem SIMD has faced with an ever increasing instruction set size and complexity as data types grow from 32 bits to 64, 128, and beyond. The Vector ISA does not face this issue, as all instructions are effectively the same, but the Vector-Length and Configuration registers hold the type and size of data to be operating. It supports the base 64-bit instruction set of RISC-V ISA, but for vector registers. The maximum vector length and supported extension instructions are left up as an implementation detail. These key details give vector architectures as a platform for great code reuse, backwards portability, and a stable target for code optimization libraries. Vector processors also hold the benefit of providing ultra-low power efficiency \cite{vecppowerinproceedings} and being well suited for accelerating parallel workloads. 

Based on this comparison and operating under the domain of resource and power constrained embedded SoCs, the vector architecture was chosen as the optimal platform potential for optimizing the performance of image pre-processing and CPU fallback layers that the NVDLA compiler and hardware architecture could not support. If successful, this should pave the way for more exploration and optimization support for this architecture across more modern CNNs, frameworks, and DLA architectures. 
\begin{table}[h]     \centering     \caption{Solution Space For Image Processing Hardware}     \label{tab:imphw}     \small    \setlength{\tabcolsep}{6pt} 
\begin{tabular}{|p{1cm}| p{1cm}| p{2.5cm}|p{2.5cm}| p{3cm}|} 
\toprule         
\textbf{Type} & \textbf{Power} & \textbf{Description} & \textbf{Strengths} & \textbf{Constraints} \\     
\hline
\midrule         
\hline
CPU & High & Scalar, general core & Management & Memory bottlenecks, few cores \\  
\hline
GPU & High & Parallel cores & High performance & High power consumption/footprint \\   
\hline
FPGA & Medium & Configurable Logic & Flexible & Complexity \\         
\hline
ASIC & Low & Custom Logic & Fast, Low power & Fixed Function, Expensive design \\  
\hline
Vector & Ultra-low & SIMT-Like Model & Low power, small footprint & Lacking Infrastructure \\   
\hline
\bottomrule     
\end{tabular} 
\end{table}

\section{Vector Integration}
A 2048 Maximum Vector Length Hwacha was attached Vector coprocessor across the Rocket-Custom-Coprocessor Interface to expose a vector-fetch programming model. This interface passes custum xHwacha RISC-V extension instructions to invoke function/procedure level independent execution on Hwacha. This becomes a tightly-coupled coprocessor to the Rocket tile which has direct access to the L2 Data cache hierarchy. The microarchitecture of Hwacha includes a pipeline of Vector memory (VMU) and processing units (VPU) with support for built in floating point computation.

\begin{figure}[h]
  \centering
  \includegraphics[width=\linewidth]{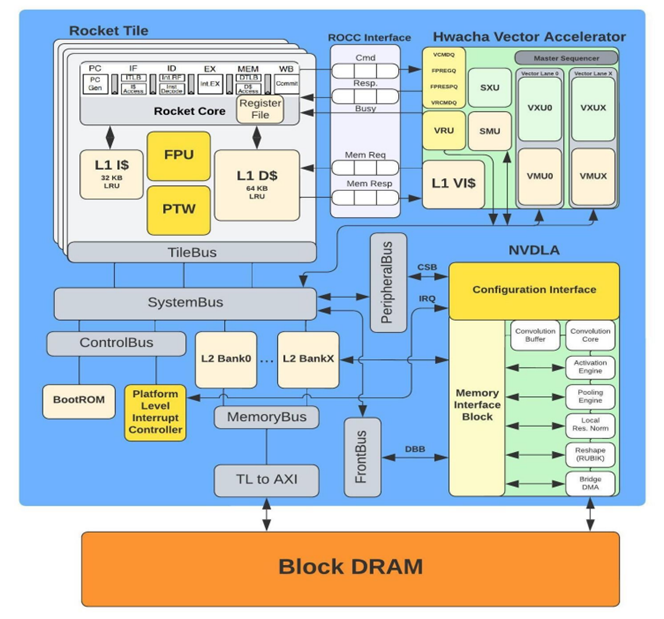}
  \caption{Hwacha Integration into Rocket-NVDLA System}
  \Description{Hwacha Integration into Rocket-NVDLA System}
\end{figure}

\subsection{Vector Mapping CPU Fallback Operations}

\begin{algorithm}
\caption{Convert FD to NCHW}
\label{algo:fd_to_nchw}
\begin{algorithmic}
\Function{nconvert\_fd\_to\_nchw}{$\text{in}, w, h, c, \text{out}$}
    \State $\text{line\_stride} \gets w \cdot 32$
    \State $\text{surface\_stride} \gets \text{line\_stride} \cdot h$
    \For{$i \gets 0$ to $c$}
        \For{$j \gets 0$ to $h$}
            \For{$k \gets 0$ to $w$}
                \State $\text{surface\_index} \gets \frac{i}{32}$
                \State $\text{out}[w \cdot h \cdot i + w \cdot j + k] \gets \text{in}[\text{surface\_stride} \cdot \text{surface\_index} + \text{line\_stride} \cdot j + 32 \cdot k + i\%32]$
            \EndFor
        \EndFor
    \EndFor
\EndFunction
\end{algorithmic}
\end{algorithm}
To demonstrate the process of vector mapping towards a RVV target, we have included the commonly used function within the conversion fallback layer: FD-to-NCHW. After a convolution-maxpooling layer, the tensor is represented as a feature-map. To advance to the next layer, the tensor must be reformated to the NCHW (number:batch, channel, width, height) format. This is an expensive operation on CPU because of the large load/stores through the cache hierarchy with little memory reuse. 
Here, the vector-mapped version of of the feature-mapped conversion code configures the vector configuration register to allocate only one vector register (to fit all elements into one). The line and surface stride is hoisted out of the loop and the surface index is found for each channel iteration. The offsets for input and output buffers are computed within the first inner loop. Within the vectorized inner loop, the vector length (the lesser of Max Vector Length and remaining pixels), the input/output offsets are moved to vector registers, as well as the line stride with 32 bit alignment. The Hwacha PC is set to the vcvt\_fd\_to\_nchw procedure. The vector fetch instruction invokes the vector accelerator to begin execution. It is important to note a fence instruction is required after this process, to prevent any race conditions or synchronization issues between the CPU and cache hierarchy. 


The entire load-store operation of the feature-depth conversion to NCHW is hoisted over to the Hwacha accerator. Its Vector Memory unit has direct access to the L2 data cache and performs a unit strided load of a block of words into the vector register 0 starting from the offset passed from the Rocket Core over the RoCC interface. It then stores the contents of the vector register at the destination address and halts. 
\subsection{Integrating Hwacha Vector Coprocessor}
All conversion layer subprocedures, and image preprocessing functions were mapped to the Hwacha coprocessor using the C-assembly directives and RISC-V binary toolchain. They were tested for functional correctness to the original DarkNet version with the Spike simulator. Initial validation on the target SoC was performed function-wise with a compiled model of the Chisel elaboration on Verilator C++ accelerated simulator. Final metrics for speedup and memory performance were taken from FireSim memory accurate FPGA accelerated simulation with a system clock speed of 100Mhz.
\subsection{Memory Performance }
Taking into account that Hwacha is streaming blocks of data from the L2 cache, there was a large bottleneck found in cache efficiency during memory accurate system simulation using FireSim. On average, there were 82.3 clock cycles in idling for cache to service Hwacha's bulk requests. This followed the fact that there was very little data reuse in the vector-mapped execution, very similar to the memory access behavior found in CNN execution on NVDLA. Since in the vector-mapped code, the address offsets were computed in the outer most loops and known in advance to Hwacha vector loads and stores, simple prefetching instructions were injected before and after each iteration, to give the cache time to begin filling its blocks. This served the linear-strided access patterns of most operations very well, incurring a speedup of approximately 3x compared to without the prefetching implementation. Hardware prefetchers integrated into the cache hierarchy are expected to further mitigate this memory squeeze.

\begin{figure*}[ht]\label{fig:cache}
    \includegraphics[width=.5\textwidth]{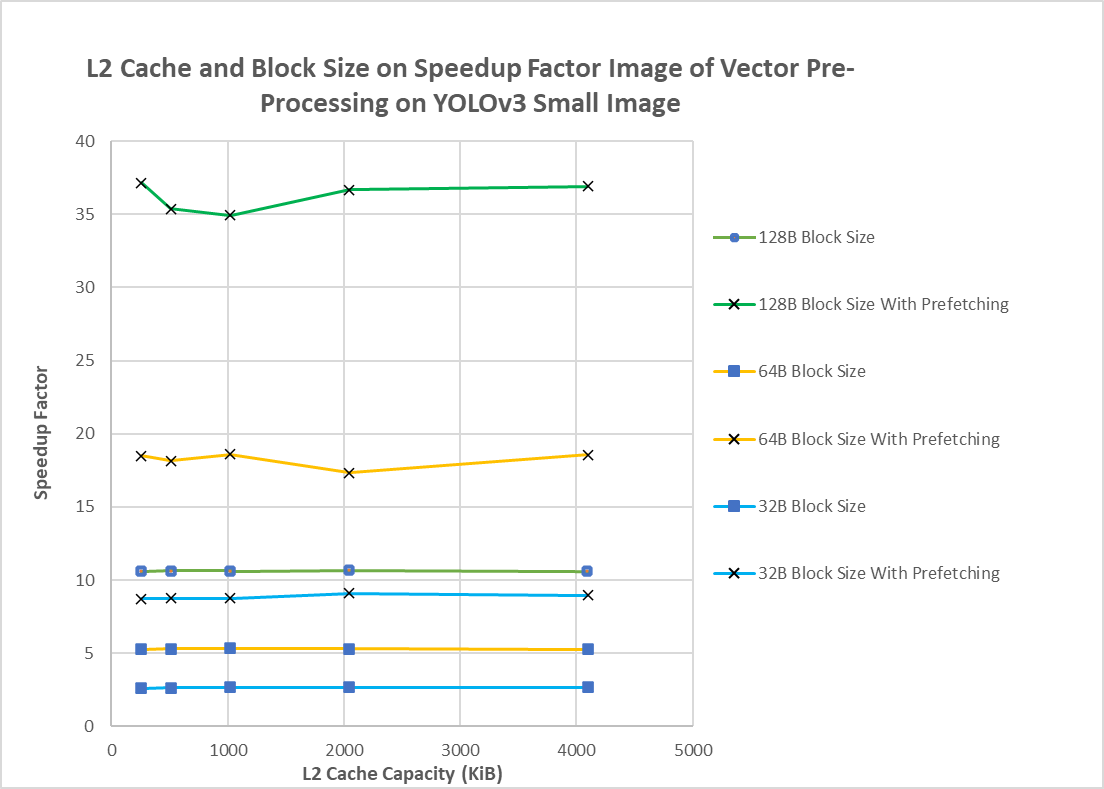}\hfill
    \includegraphics[width=.5\textwidth]{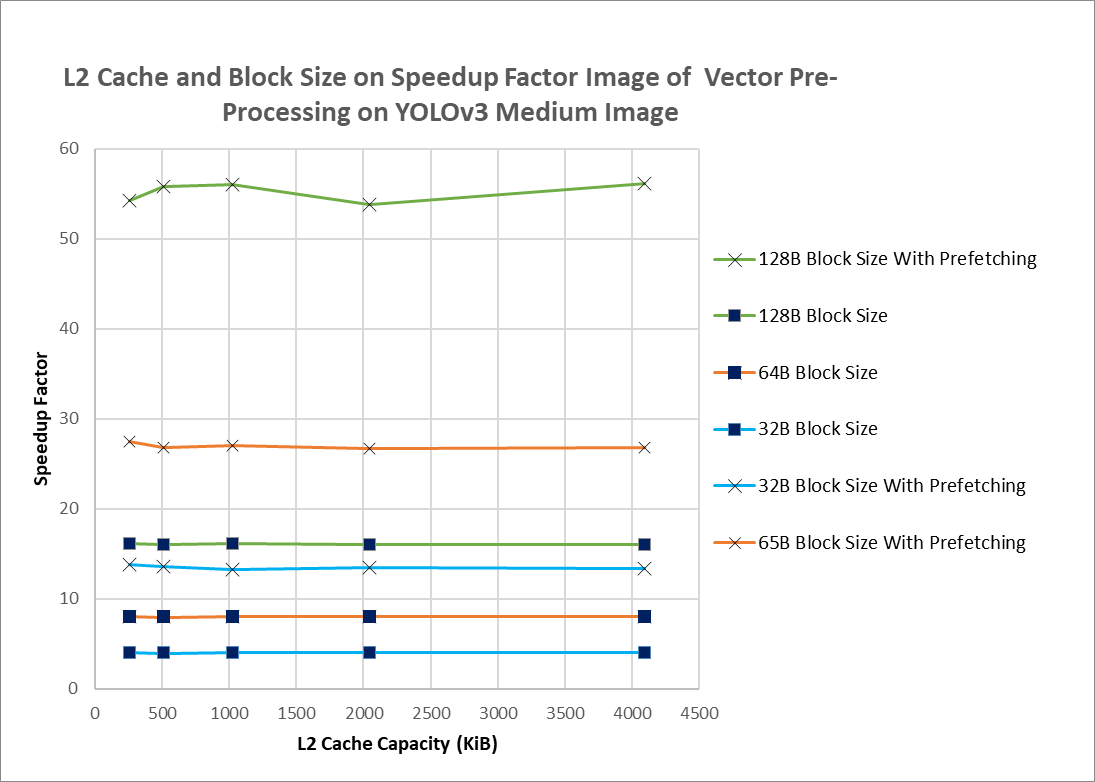}\hfill
    \includegraphics[width=.5\textwidth]{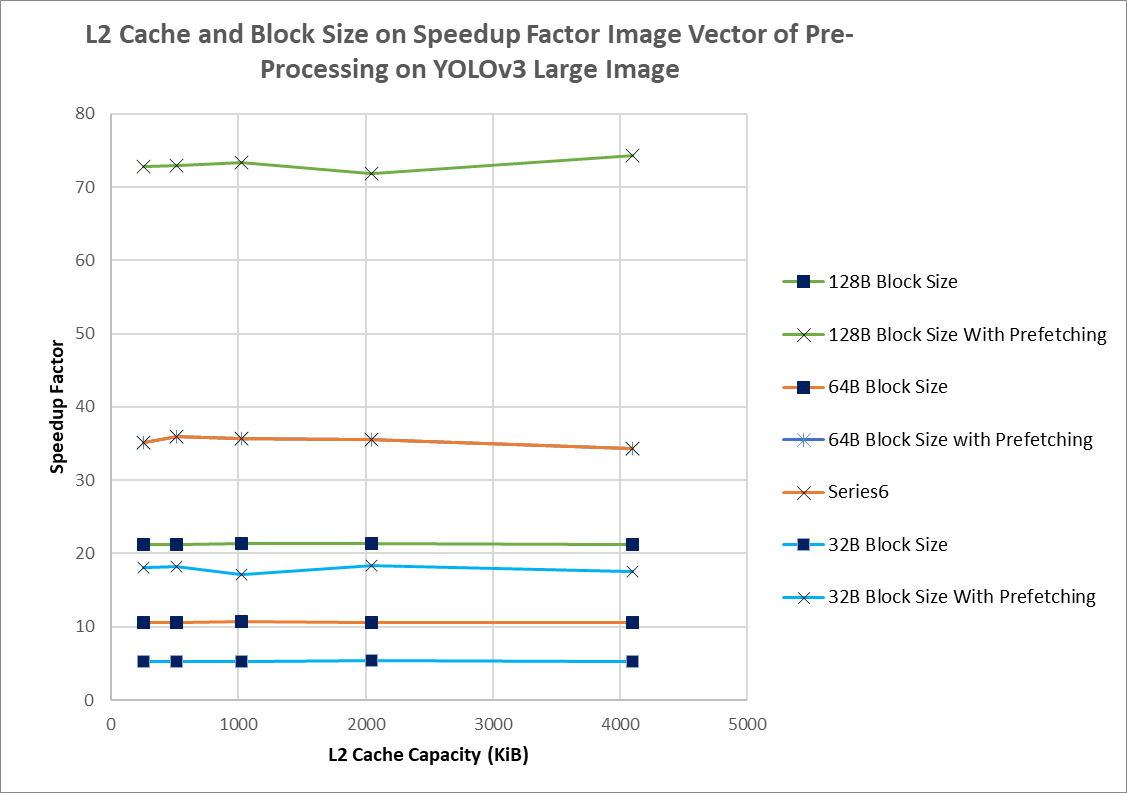}\hfill
\end{figure*}
 
\subsection{Results}
After implementing prefetching in the vector mapped implementation of the conversion layer and image preprocessing pipeline, there was between 3-72x  speedup compared to scalar, single threaded CPU implementation, for which vectorization was possible. Some functions were chosen not to be vector-mapped, such as Non-max suppression. This included many many branching and control flow operations, and despite being a slow CPU post-processing function, did not exhibit much potential for improvement. YOLO cost calculation and upsampling metrics will remain as future work, although a few subprocedures have been implemented. We expect to see a similar improvement there. We propose auto-vectorization or vector-aware technology being integrated into the standard RISC-V toolchain will ease the burden on software designers from performing handwritten vector-mapping.  
\begin{figure*}[ht]
    \includegraphics[width=.5\textwidth]{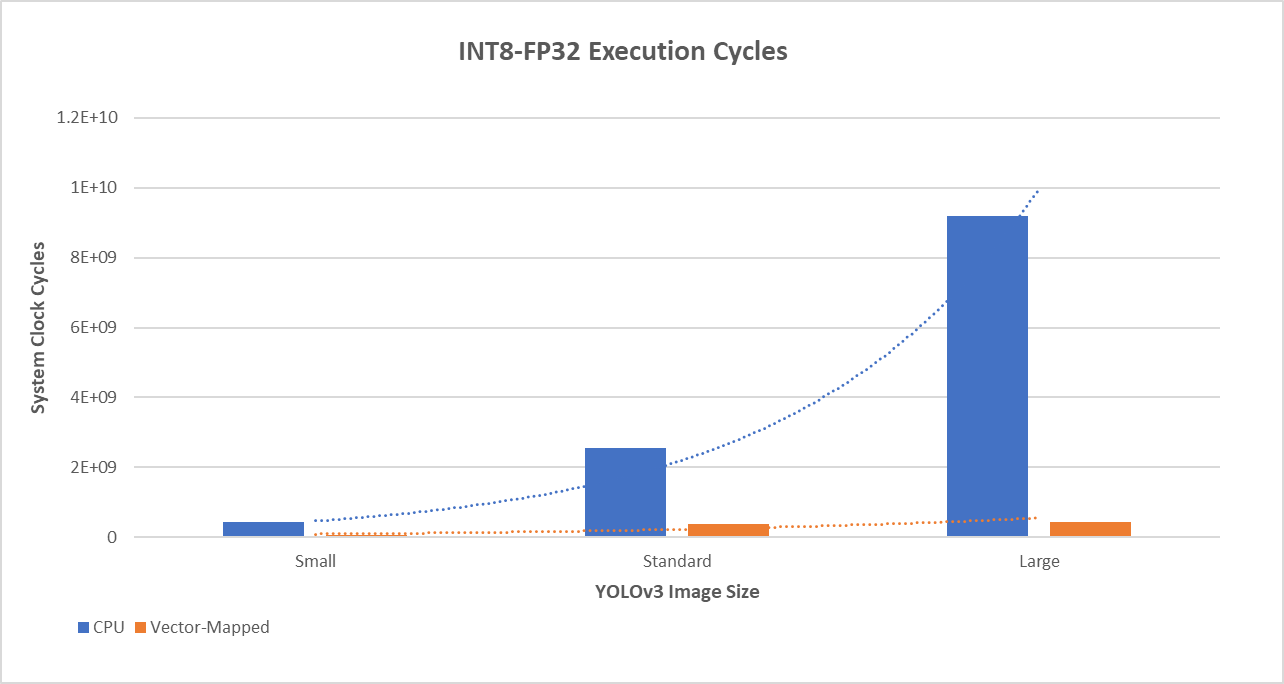}\hfill
    \includegraphics[width=.5\textwidth]{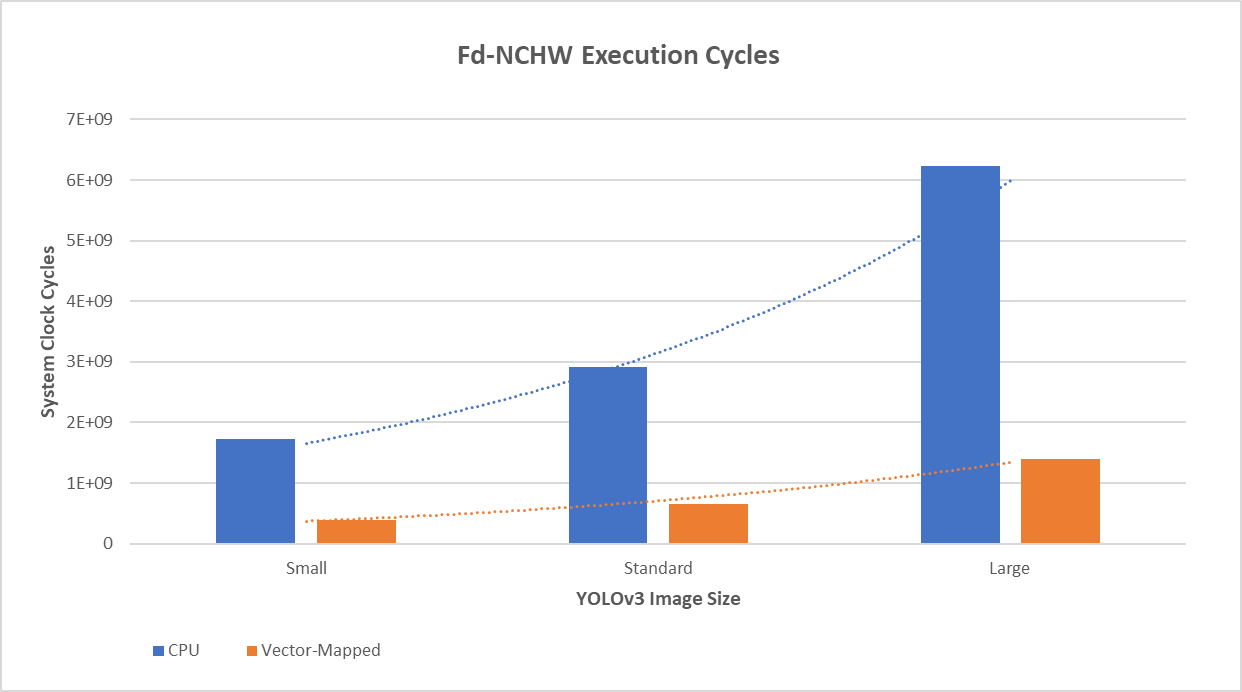}\par
    \includegraphics[width=.5\textwidth]{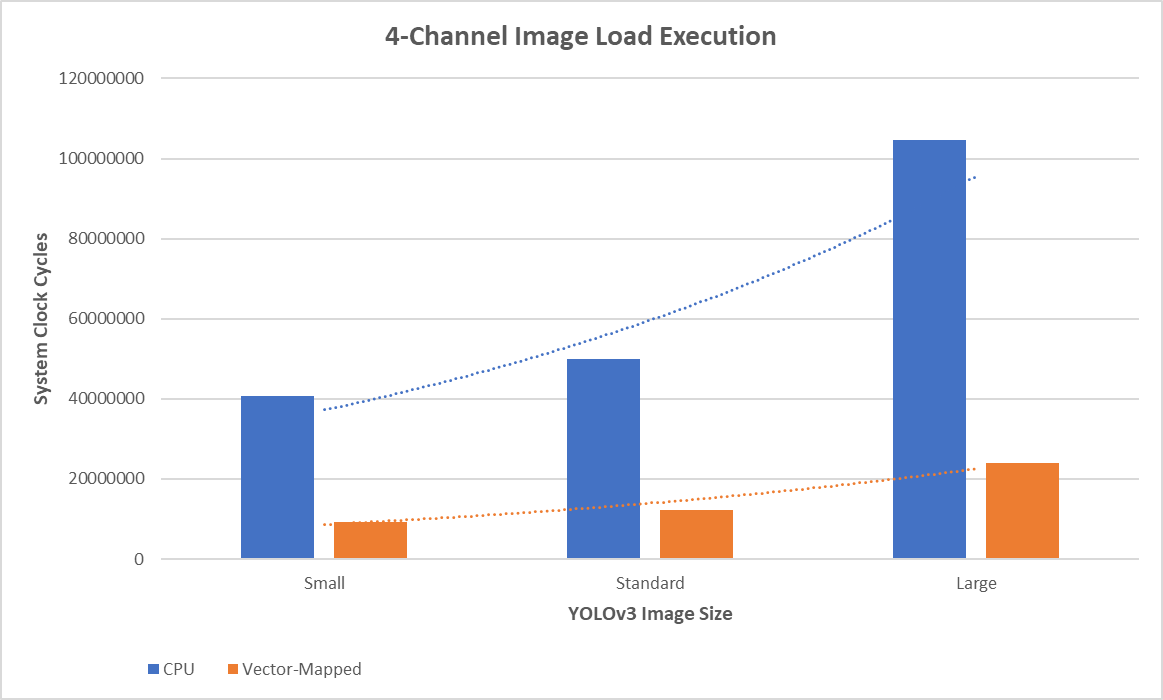}\hfill
    \includegraphics[width=.5\textwidth]{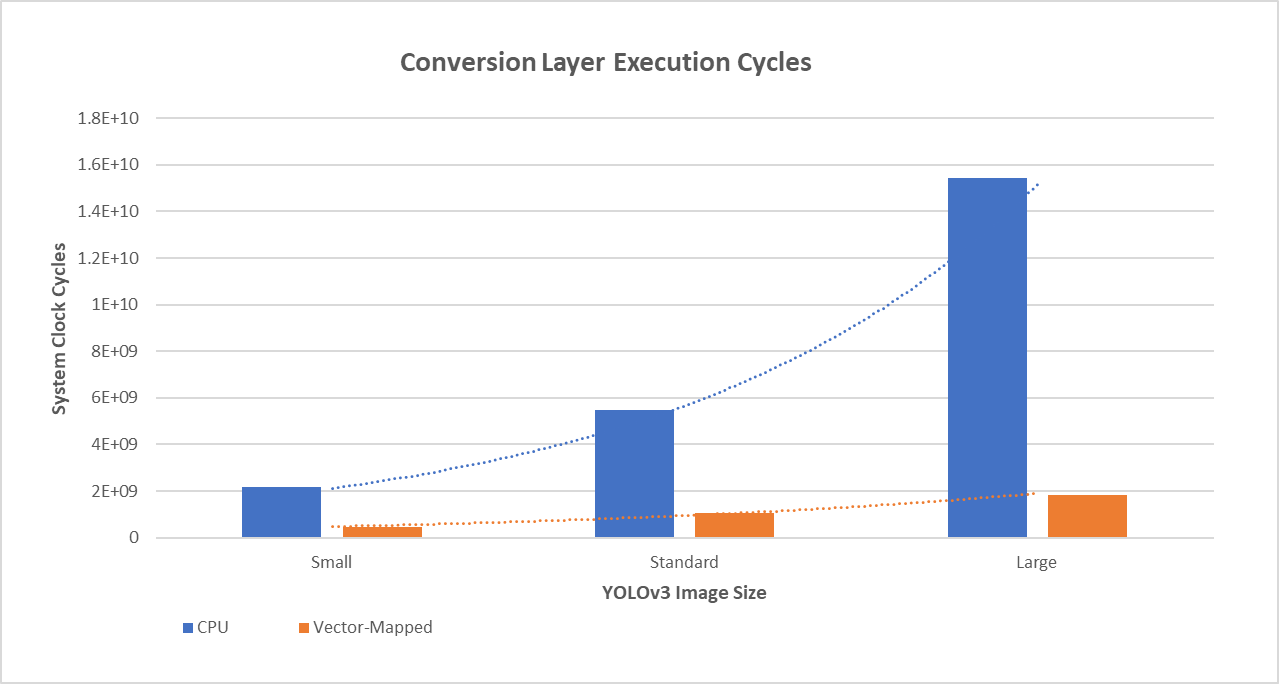}\par
\end{figure*}
\begin{table}[ht] 
    \centering 
    \caption{Hwacha Speedup With Non-Blocking Cache Prefetching} 
    \label{tab:hpmcounter} 
    \begin{tabular}{ccc} 
        \toprule 
        \textbf{Workload} & \textbf{Image Size} & \textbf{Speedup} \\ 
        \midrule  
        Converter & Small & 4.601x \\ 
        Converter & Medium & 8.638x \\ 
        Converter & Large &  9.934x\\ 
        Total & Small  & 2.260x \\
        Total & Medium  & 3.003x \\ 
        Total & Large  & 3.668x \\

        \bottomrule 
    \end{tabular} 
\end{table} 
\section{Related Work}
RISC-V's revival of the vector architecture has paved the way for recent exploration into its flexible and performant feasibility to execute or support machine learning and deep learning. Al Assir et. al \cite{arrowDBLP:journals/corr/abs-2107-07169} shows that one implementation of the RVV v0.9 extension achieves up to 78x faster performance on matrix/vector benchmarks fundamental to machine learning with between 20\% and 99\% less energy. On the MicroBlaze FPGA, LUT usage was less than 500, whereas our configuration of Hwacha used 103290 on the VCU188 platform. Another implementation of RISC-V vector processor is integrated into the Rocket chip system for end-to-end in-pipeline support for machine learning and DNN inference such as YOLO-tiny, DenseNet and ResNet-50 \cite{harvard1594785}. It highlights the special attention required for the entire dataflow: pre-processing, inference, and post-processing. They intend to deploy the maximal amount of the inference execution to the vector coprocessor.
That work's main takeaway is the ease of integration into RISC-V SoC's and also the power of extensibility in implementation of the vector specification. It is still awaiting concrete power and implementation performance results. An AI Vector accelerator has also been developed for complete neural network inference \cite{AIVectOn&nbsp}. This developed and demonstrated the efficiency and code portability and compatibility with ingesting DNN models (defined in TensorFlow Lite) and outputting efficient code generation  for the target. By creating a RVV implementation that optimizes for convolution, relu, addition, and multiply, the software-hardware codesign acheived respectable results of 7.3x speedup compared to a 32bit RISC-V CPU based on the tinyMLPerf \cite{TinyMLDBLP:journals/corr/abs-2106-07597} benchmark. It is awaiting verification beyond a Verilator model, and expects an FPGA evaluation in future work. These works have shown early demonstrations that the vector coprocessor is a flexible target for optimizing around the complete dataflow of deep learning. However, it is clear that custom DLAs with streaming, neuromorphic, or dedicated operation architectures will continue to outperform vector processors for the bulk of these tasks \cite{lhwDBLP:journals/corr/abs-2104-09252}, therefore it is reasonable to assume that most compiler, software and system level designs will target support and deployment of most CNN operations for DLAs, not vector units. With this in mind, our work focuses on improving the CPU fallback and pre-processing performance alone, targeting an embedded design that is centered around a DLA, but providing a balanced memory and execution footprint with the help of a vector fallback unit.
\section{Conclusion}
This work successfully provided an analysis of software and hardware bottlenecks in resource constrained embedded systems targeting demanding streaming CNN inference with the design goal of high performance per watt. The YOLOv3 model was investigated in its implementation and deployment on the Nvidia Deep Learning Accelerator and many CPU fallback operations were discovered to be be fairly simple. This demonstrated the immense need for deep learning compiler technology to have suitable hardware-aware support for complex modern CNN models with varying precision and layer types, with efficient fallback targets for unsupported operations and pre-processing. We proposed the Hwacha \cite{Hwacha-Micro} vector accelerator as a low-power, high performance fallback target that takes advantage of the parallel execution nature that was discovered in much of the work. The integration with the RISC-V cache hierarchy was tuned and further sped up with the increasing of L2 cache block sizes and the implementation of pre-fetching. This work served to validate the RISC-V Vector 1.0 ratification as an important step in providing a stable and extensible software platform for end-to-end CNN inference acceleration. 
\section{Acknowledgements}
We would like to thank Dr. Wafa Elmannai from Manhattan College for her valuable comments and suggestions and support for conducting this research, and the Manhattan College Jasper Summer Research Program for providing funding of this project.

\input{output.bbl}

\appendix
\section{Feature Map to NCHW Vectorization}\label{A:hwacha}
\begin{lstlisting}[caption={convert\_fd\_to\_nchw}][H] 
 // Hwacha vector implementation to convert feature depth to channel, height, width format
 // Execution in cycles = 5 + Channels * (Height * 2) + (8 * w / MAXVL)
void convert_fd_to_nchw(float* in, int w, int h, int c, float *out)
{
    // config for one word vector register and one predicate
    setvcfg(0, 1, 0, 1); 
    // number of elements in a single row
    unsigned int line_stride = w * 32; 
    // number of elements in a single surface (32 channels) of the FD input 
    unsigned int surface_stride = line_stride * h;
    for (int i = 0; i < c; ) {
        int surface_index = i / 32;
        for (int j = 0; j < h;) {
            unsigned int out_offset = (w * h * i + w * j);
            unsigned int in_offset = (surface_stride * surface_index + line_stride * j + i);
            for (int k = 0; k < w;) {
                int consumed = setvlen(w - i);
                asm volatile ("vmca va0, %0"
                                :
                                : "r" (&in[in_offset + k]));
                asm volatile ("vmca va1, %0"
                                :
                                : "r" (&out[out_offset + k]));
                asm volatile ("vmca va2, %0"
                                :
                                : "r" (32 * 4)); // stride of 128 bytes
                asm volatile ("la t0, vcvt_fd_to_nchw"
                                :
                                :
                                : "t0");
                asm volatile ("lw t1, 0(t0)");
                asm volatile ("vf 0(t0)");
                i += consumed;
            }
        }
    }
    asm volatile ("fence");
}
\end{lstlisting}

\end{document}